\documentclass[aps,amsmath,amssymb,prl,reprint]{revtex4-1}
\usepackage{graphicx,subfigure,color}

\renewcommand{\vec}{\mathbf}

\newcommand{\bul}{\hspace{-0.02in}\bullet\hspace{-0.01in}}

\newcommand{\veps}{\varepsilon}

\newcommand{\sgn}{\text{sgn}}
\newcommand{\bkp}{\mbox{\boldmath $k_{\parallel}$}}
\newcommand{\bfA}{\mathbf{A}}
\newcommand{\bfB}{\mathbf{B}}

\newcommand{\bfE}{\mathbf{E}}

\newcommand{\bfM}{\mathbf{M}}
\newcommand{\bfO}{\mathbf{O}}
\newcommand{\bfT}{\mathbf{T}}
\newcommand{\bfX}{\mathbf{X}}

\newcommand{\bfU}{\mathbf{U}}

\newcommand{\bft}{\mathbf{t}}

\newcommand{\bfa}{\mathbf{a}}
\newcommand{\bfb}{\mathbf{b}}
\newcommand{\bfc}{\mathbf{c}}
\newcommand{\bfd}{\mathbf{d}}
\newcommand{\bfg}{\mathbf{g}}
\newcommand{\bff}{\mathbf{f}}
\newcommand{\bfz}{\mathbf{z}}
\newcommand{\bfu}{\mathbf{u}}
\newcommand{\bfv}{\mathbf{v}}

\newcommand{\bfLam}{\boldsymbol{\Lambda}}
\newcommand{\diag}{\text{\bf diag}}

\begin{document}
\title{Topological phase of the interlayer exchange coupling with application to magnetic switching}

\author{A. Umerski}
\affiliation{Department of Mathematics and Statistics, Open University, Milton Keynes MK7 6AA,
U.K.}

\date{\today}

\begin{abstract}
We show, theoretically, that the phase of the interlayer exchange coupling (IEC) undergoes a topological change of approximately $2\pi$ as the chemical potential of the ferromagnetic (FM) lead moves across a hybridization gap (HG). The effect is largely independent of the detailed parameters of the system, in particular the width of the gap. 
The implication is that for a narrow gap, a small perturbation in the chemical potential of the lead can give a sign reversal of the exchange coupling.
This offers the possibility of controlling magnetization switching in spintronic devices such as MRAM, with little power consumption. 
Furthermore we believe that this effect has already been indirectly observed, in existing measurements of the IEC as a function of temperature and of doping of the leads.
\end{abstract}

\maketitle
%


A principal contender for the next generation of spintronic memory devices is the magnetic random access memory (MRAM) \cite{Butler04,Akerman2005}.
The basic component of this device is a tunnel junction consisting of two ferromagnetic (FM) leads separated by a few atomic layers of crystalline magnesium oxide.
The magnetic moments of the FM leads can either be parallel (P) or antiparallel (AP) and it is found that the junction's electrical resistance can be over 100 times higher in the AP state, which makes it ideal for storing and reading information.

However the development of this approach has been hampered by the difficulty in finding an efficient mechanism to invoke writing i.e.  switching between the P and AP states.
To date, the majority of research has focussed on using a spin-polarised current to effect switching via spin-transfer torque (STT-MRAM) \cite{Slonczewski96,Slaughter12}.
However a high current density is required, which is an obstacle to creating energy-efficient nano-scale devices.

An alternative approach is to use the oscillatory interlayer exchange coupling (IEC), in which two FM leads are separated by a few atomic layers of a non-magnetic (NM) metallic spacer \cite{Parkin90}. 
Combining such a device with a tunnel junction might provide an efficient way of effecting switching, particularly if the state of the IEC system can be switched from P to AP by applying a voltage or electric field \cite{Matsukura15, Newhouse17}. 
Such a system could be switched using little power so would be both energy-efficient and scalable.

Here we concentrate on the manipulation of the potentials in the lead of the IEC device. This could be achieved, for example, by attaching a semiconductor to a conventional FM lead and applying a bias voltage to manipulate the Schottky barrier height formed at the interface \cite{Suzuki05}, or by applying a bias voltage to a ferromagnetic semiconducting lead \cite{Matsukura15} (other plausible mechanisms for switching of IEC have also been investigated \cite{Bader99, Tsymbal10, Fechner12, Newhouse17}).
With this goal in mind we investigate theoretically the influence of the chemical potential in the FM lead on the phase of the IEC and we find a surprising result: when the Fermi-level of the system lies in a HG of the FM lead, there is a topological change of phase of $\approx2\pi$ in a component of the IEC as the chemical potential moves across the gap. The effect is largely independent of the detailed parameters of the system, in particular the width of the gap.
The implication is that if the HG is narrow, a very small perturbation in the potential can give a sign reversal of the exchange coupling.


We begin by following the discussion given in Ref. \cite{Mathon97} and 
consider a FM/NM/FM trilayer, with a NM metallic spacer of thickness $n$ atomic planes and semi-infinite FM leads.  
 The exchange coupling per surface atom is
given in terms of the thermodynamic potentials $\Omega^{\sigma}_{\rm P/AP}$ for electrons of spin $\sigma$, in configuration P/AP, by
\begin{equation}
J(n) = \Omega_{\rm P}^{\uparrow}(n) + \Omega_{\rm P}^{\downarrow}(n) - \Omega_{\rm AP}^{\uparrow}(n) - \Omega_{\rm AP}^{\downarrow}(n).
\label{eq:IEC}
\end{equation}
We call the $\Omega$'s the `IEC components', as they can differ from the true thermodynamic potentials by terms which cancel in the overall sum of the IEC.

We assume that the system grows epitaxially with in-plane translational symmetry, so we label states by a planar index $n$ and wave vector $\vec{k}_{\parallel}$ parallel to the layers.
Then it can be shown that if the spacer has a single Fermi-surface (FS) branch, then the IEC components at temperature $T$ are given by the following asymptotically exact formula, which gives excellent agreement with fully realistic calculations \cite{Mathon95,Mathon97,Costa97} (from here on we drop the P, AP, $\sigma$ labels and concentrate on a single IEC component).
\begin{equation}
\Omega(n) \propto  \frac{T}{n} 
 \sum_{{\bf k}_{\parallel}^0} \sum_{s} \Re\Biggl( \frac{A(k) \,|c_{s}| e^{i(2ns k+\psi_s)}}{s\, \sinh(\pi k_B T[2ns k^{\prime} + \psi_{s}^{\prime}])}\Biggr) {\Biggr|}_{\veps=\mu}.
\label{eq:SPA}
\end{equation} 
More general formulas exist when the spacer FS has multiple branches but here we concentrate on the single branch case.
All quantities on the right-hand side of equation~(\ref{eq:SPA}) are evaluated at the common Fermi level $\veps = \mu$ and
the first sum is over the stationary points ${\bf k}_{\parallel}^0$ of the spacer FS. 
$A(k)$ is a real constant dependent on the curvature of the spacer FS.
$k$ is the Fermi wave vector of the spacer in the growth direction (perpendicular to $\bkp$), and 
$k^{\prime}=dk/d\veps$ is the inverse FS velocity.
Finally the second sum is over the Fourier coefficients $c_s=|c_s|e^{i\psi_s}$ (s=1,2,\ldots) of the electron number density which can be shown to be periodic with period $\pi/k$.  
$\psi_s$ is the phase of the Fourier coefficient $c_s$ and $\psi_{s}^{\prime}=d\psi_s/d\veps$.
 
Note that almost the entire expression on the right hand side depends only on the NM spacer. The only terms dependent on the FM lead come in through the Fourier coefficients.  
Hence these are the only terms which differ between $\Omega_{\rm P}^{\uparrow}$, $\Omega_{\rm P}^{\downarrow}$, $\Omega_{\rm AP}^{\uparrow}$ and $\Omega_{\rm AP}^{\downarrow}$.
We also note that $\Omega(n)$ and hence $J(n)$ oscillates and decays with $n$, with a $\pi/k$ oscillation length as in RKKY theory \cite{Bruno91}.  Furthermore the phase of the oscillation depends entirely on $\psi_s$,
and so if we are to understand the phase of the IEC we need to understand the phase of the Fourier coefficient.
 
To calculate the Fourier coefficients we extend the theory of matrix M\"{o}bius transformations (MT) introduced in Ref. \cite{Umerski97}.
This is a generalisation of the well-known conformal transformation which maps circles and lines into circles and lines in the complex plane, defined by
\[
\bfA\bul\bfz \equiv (\bfa\bfz+\bfb)(\bfc\bfz+\bfd)^{-1}, \quad \text{where} \quad \bfA = 
\begin{pmatrix}\bfa&\bfb\\\bfc&\bfd\end{pmatrix}.
\]
Here $\bfa$, $\bfb$, $\bfc$, $\bfd$ and $\bfz$ are $N \times N$ matrices and  
it is straightforward to show that this transformation is associative with respect to multiplication
$\bfA\bul(\bfB\bul\bfz) = (\bfA\bfB)\bul\bfz$.  
This construct now quite widely used to calculate surface Green's functions (SGF) and is particularly useful for understanding the properties of layered systems. 
In this context, given a uniform system with Hamiltonian defined by the $N \times N$ on-site matrix $\bfu$ and hopping matrix $\bft$, then we define the MT matrix $\bfX$ by
\[\bfX = \begin{pmatrix} \mathbf{0}&\bft^{-1}\cr -\bft^{\dag}& \bfv\bft^{-1} \end{pmatrix},
\]
where $\bfv \equiv \veps - \bfu$ and $\veps$ is the energy and where $\mathbf{0}$ is the $N \times N$ matrix of zeros.
Next we define its eigenvalue matrix $\bfLam$ and eigenvector matrix $\bfO$ by
\[
{\bf O}^{-1} {\bf X} {\bf O} = {\bf \Lambda}
= \diag(\lambda_1,\lambda_2, ... ,\lambda_{2N})\]
and we order the eigenvalues so that $|\lambda_1| < |\lambda_2| < ... < |\lambda_{2N}|$ (introducing a small imaginary part to the energy $\veps \rightarrow \veps + i\delta$ to do this uniquely).
Denoting the MT matrix composed of lead potentials by $\bfX_L$ and computing its eigenvector matrix $\bfO_L$, 
gives a simple expression for the semi-infinite left-hand lead SGF \cite{Umerski97}:
\begin{equation}
\bfg_{0}=\bfO_{L}\bul\mathbf{0}.
\label{eq:g0}
\end{equation}
Similar expressions are obtained for the right-hand lead SGF $\bfg_{n+1}$  by replacing $\bft$ by $\bft^{\dagger}$ throughout.

Then it can be shown that \cite{Umerski17} for the case where the spacer has a single Fermi-surface branch, the Fourier coefficients of the number density are given by 
\begin{equation}
c_1=e^{2ik}(\bff_0)_{N,1}(\bff_{n+1})_{1,N}\ ,\  c_s=\frac{c_1^s}{s}\ , \ s=1,2,\ldots\ ,
\label{eq:c1}
\end{equation}
where the sign of the spacer Fermi wave-vector $k$ is chosen such that $k'>0$. 
Here
$\bff_0=\bfO_S^{-1}\bul\bfg_0$ and $\bff_{n+1}=\bfO_S^{-1}\bul\bfg_{n+1}$ are $N \times N$ matrices given by the MT of the left and right lead SGFs respectively, and $\bfO_S$ is the matrix of eigenvectors derived from the MT matrix with spacer potentials.
Note that $\bff_0$ ($\bff_{n+1}$) depends only on the left (right) lead and the spacer.

We are particularly interested in how the phase of the IEC, $\psi_s$, changes as we vary the potential in for example the left lead:
clearly this is determined by the phase change of $(\bff_0)_{N,1}$.
For the 1-band model it can be shown that \cite{Umerski17} as the potential in one lead varies from $-\infty$ to $+\infty$, 
the phase of the IEC components are topologically constrained to change by $0$ or a full $2\pi$ radians, depending on whether $|t/T|<1$ or $|t/T|>1$ respectively (where $t$ ($T$) is the spacer (lead) hopping). 
However this happens over a large energy range so is not of great practical importance, although the rate of change of phase $\psi_s'=d\psi_s/d\veps$ can be quite large very near the band edge.
Another feature which emerges is that the magnitude of $c_1$ depends on the degree of confinement of the quantum well states in the spacer and has a maximum value of $|c_1|=1$ when both leads are insulating.
This latter result is in agreement with previous findings \cite{Mathon97} which showed that the IEC is largest when quantum well states are fully confined.

The fact that the 1-band model shows that there are topological constraints on the phase of the IEC components, and their magnitudes  are largest when quantum well states are fully confined suggests that it might be fruitful to study the IEC in the case where the chemical potential of the lead moves through a HG. 
To do this we first study a 2-band model, in which (for one of the spins) the lead on-site potential $\bfU$ and hopping $\bfT$ are given by
 \[\bfU=\begin{pmatrix} 0&T\\T&U \end{pmatrix} \quad,\quad \bfT=\begin{pmatrix} \frac12 &S\\ S& -\frac12 \end{pmatrix} \quad,\quad T\ge 0.\]
 (It is possible to analyse more general models, but the resulting expressions become algebraically complicated and no new insights are gained.) 
Solving the dispersion relation $\bfE(K)=\bfU+2\bfT\cos K$, shows that for $\frac12 U \le 4S^2+2ST+1$ this model has a HG for $\veps_B < \veps < \veps_T$ where $\veps_{T/B}= \frac12(U \pm W)$, and the minimum gap width $W=2|T+US|/\sqrt{1+4S^2}$ occurs for $0<K<\pi$.
(For $\frac12 U > 4S^2+2ST+1$ the minimum gap width occurs at $K=0$ or $\pi$ and can also be analysed using the methods developed below \cite{Umerski17}.)

For the NM spacer, we consider a particularly simple 2-band model (two independent 1-band models) with on-site potential and hopping
\[\bfu=\begin{pmatrix} u_1&0\\0&u_2 \end{pmatrix} \quad,\quad \bft=\begin{pmatrix} t &0\\0& t \end{pmatrix}, \]
and we choose our potentials such that only one band crosses the HG: $u_1-2|t|<\veps_T$, $u_1+2|t|>\veps_B$, $u_2 \gg \veps_T+2|t|$.
Then the energy bands are $E_i(k)=u_i+2t\cos k$, and by solving these for $k$ we obtain ${k}_i=\cos^{-1} (v_i/2t)$, where $v_i=\veps-u_i$.

For energies in the HG, $k_1$ is real and $k_2$ is pure imaginary, so  we define $k=\pm{k}_1$ and $\kappa=\pm i{k}_2$ to be both real with signs chosen such that $\kappa>0$ and $k'>0$.
Then from equation~(\ref{eq:c1}) only the $(2,1)$ component of $\bff_0$ contributes to the Fourier coefficient and can be shown to be \cite{Umerski17}
\begin{equation}
(\bff_0)_{21}=(\bfO^{-1}\bul\bfg_0)_{21}=\frac{\det \bfM}{(\det \bfM)^*},
\label{eq:f0}
\end{equation}
where
\begin{equation}
\det \bfM = 1-e^{-ik}t(\bfg_0)_{11}-e^{-\kappa}t(\bfg_0)_{22} + e^{-ik}e^{-\kappa}t^2 \det \bfg_0.
\label{eq:detM}
\end{equation}
So, $(\bff_0)_{21}$ has modulus unity, and we can write $(\bff_0)_{21}=e^{2i\psi}$ where $\psi$ is the phase of $\det \bfM$.
Thus we wish to evaluate the change in phase of $\det \bfM$ as the chemical potential in the lead crosses the HG.  
This is equivalent to studying $\det \bfM$ as we vary the energy $\veps$ in $\bfg_0(\veps)$ from $\veps=\veps_T$ to $\veps=\veps_B$ (the spacer Fermi wave-vectors $k$ and $\kappa$ remain frozen).

\begin{figure}[tp]
\begin{center}
\includegraphics[width=1.0\linewidth]{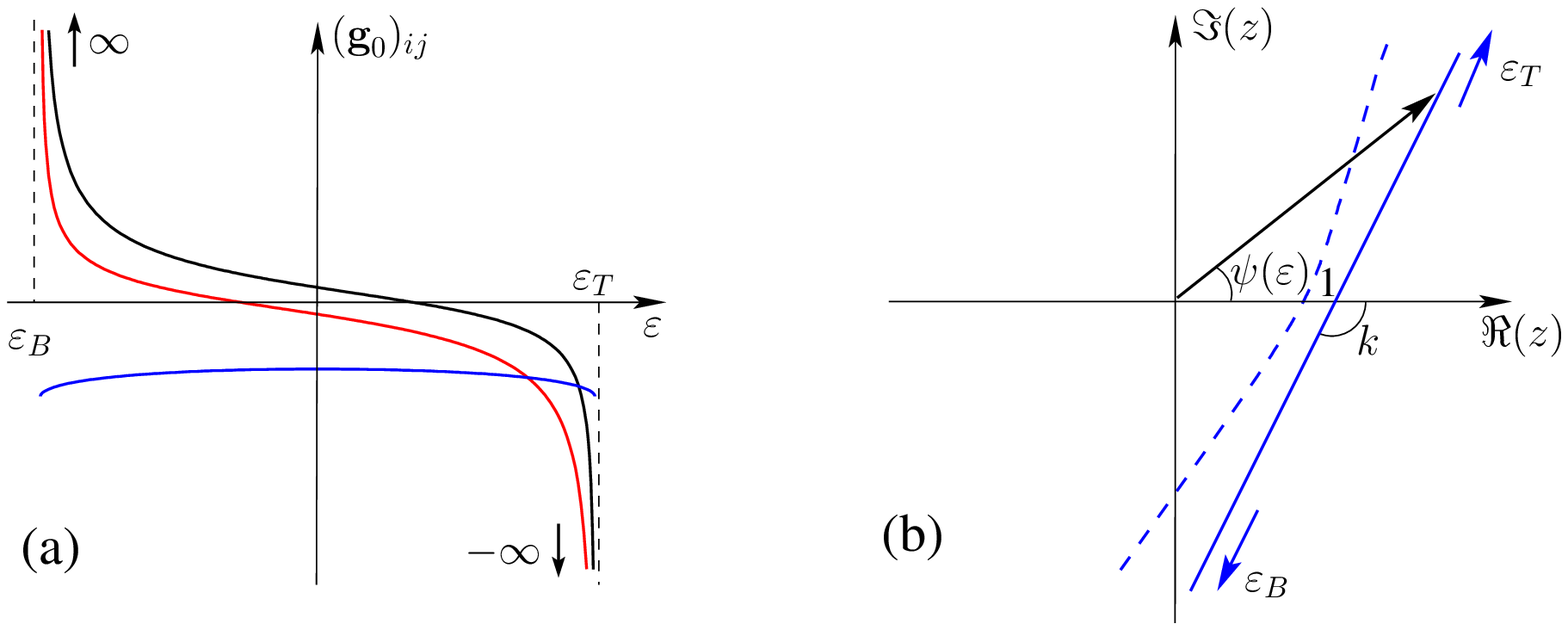}
\end{center}
\vspace{-0.2in}
\caption{\footnotesize (a) Typical paths of $(\bfg_0)_{11}$ (black), $(\bfg_0)_{22}$ (red), $\det \bfg_0$ (blue),  for $\veps_B<\veps<\veps_T$. 
(b) Typical paths of $\det \bfM$  for $\veps_B<\veps<\veps_T$. Solid line for $\kappa \gg 1$, dashed curve for moderately large $\kappa$.}
\label{fig:g11}
\end{figure}

We calculate the SGF $\bfg_0$ of the lead analytically using equation~(\ref{eq:g0}). 
We leave the details for a future paper and present the result that in the HG
\[
\bfg_0(\varepsilon)=\frac{2e^{2\Im (K)}}{(1+4S^2)\Im(x)} \Im \left [ e^{iK} (1 + 2Sx)
\begin{pmatrix} 1/{x^*} & 1 \\ 1 & x^*\end{pmatrix} \right],
\]
where
\[x(\veps)=\frac{\veps-\cos K}{T+2S\cos K}\ ,\]
and $K$ is the exponent of the complex eigenvalue $\lambda_3$ of the MT matrix for the lead: $\lambda_3=e^{iK}$.
Note that $g_0$ is real for energies in the gap and it can be shown that $(\bfg_0)_{ii} \rightarrow -\infty$ ($+\infty$) as $\veps \rightarrow \veps_T$ ($\veps \rightarrow \veps_B$), but $\det \bfg_0$ remains finite throughout (see Fig.\ref{fig:g11}(a)). (These properties are likely to be generic for all systems with a HG and are the source of the $2\pi$ phase change.)
Because of this, relatively straightforward analysis further shows that as the energy $\veps$ crosses the HG, $\det \bfM$ traces out a path as depicted in Fig.\ref{fig:g11}(b), and that the overall change in phase is given by
\begin{eqnarray}
\Delta \psi &=& \psi(\veps_T)-\psi(\veps_B)\cr
&\approx& \pi + 8S \Sigma\, e^{-\kappa}\sqrt{1+4S^2}\sin k + O(e^{-2\kappa}),
\label{eq:dpsi}
\end{eqnarray}
where $\Sigma=\sgn(T+US)$ and $\sgn$ is the sign function.

From equation~(\ref{eq:c1}), $2s\Delta \psi$ must be the change in phase of $c_s$ and hence the change of phase of the IEC component $\Omega$ at a given extrema ${\bf k}_{\parallel}^0$ must be $\approx 2\Delta \psi$. (Here we ignore the $c_s$ for $s>1$ as from equation~(\ref{eq:SPA}) and (\ref{eq:c1}) they decay quickly in $s$.)
So in the limit where the second band lies far above the HG ($\kappa \gg 1$) the phase of $\Omega$ undergoes a change of $2\pi$ as the chemical potential in the lead moves across the HG, independent of the detailed parameters of the system i.e. this is a topological effect depending only on the behaviour of $\bfg$ at the HG edge.
For smaller values of $\kappa$ the change in phase can deviate from $2\pi$.
 
For the case when both leads are identical, similar results apply to $(\bff_{n+1})_{1,N}$, and hence $c_1$ will have modulus unity and as the chemical potential in both leads moves across the HG then there will be a change of $\approx 4\pi$ in the phase of $\Omega$.
Note that if potentials across the entire system can be varied using an applied bias, then there will be an additional phase change due to the change of spacer period i.e. quantum well states crossing the Fermi-level. For large spacer thickness and HG gap width this will lead to an even greater variation of $\psi$.

\begin{figure}[tp]
\begin{center}
\includegraphics[width=1.0\linewidth]{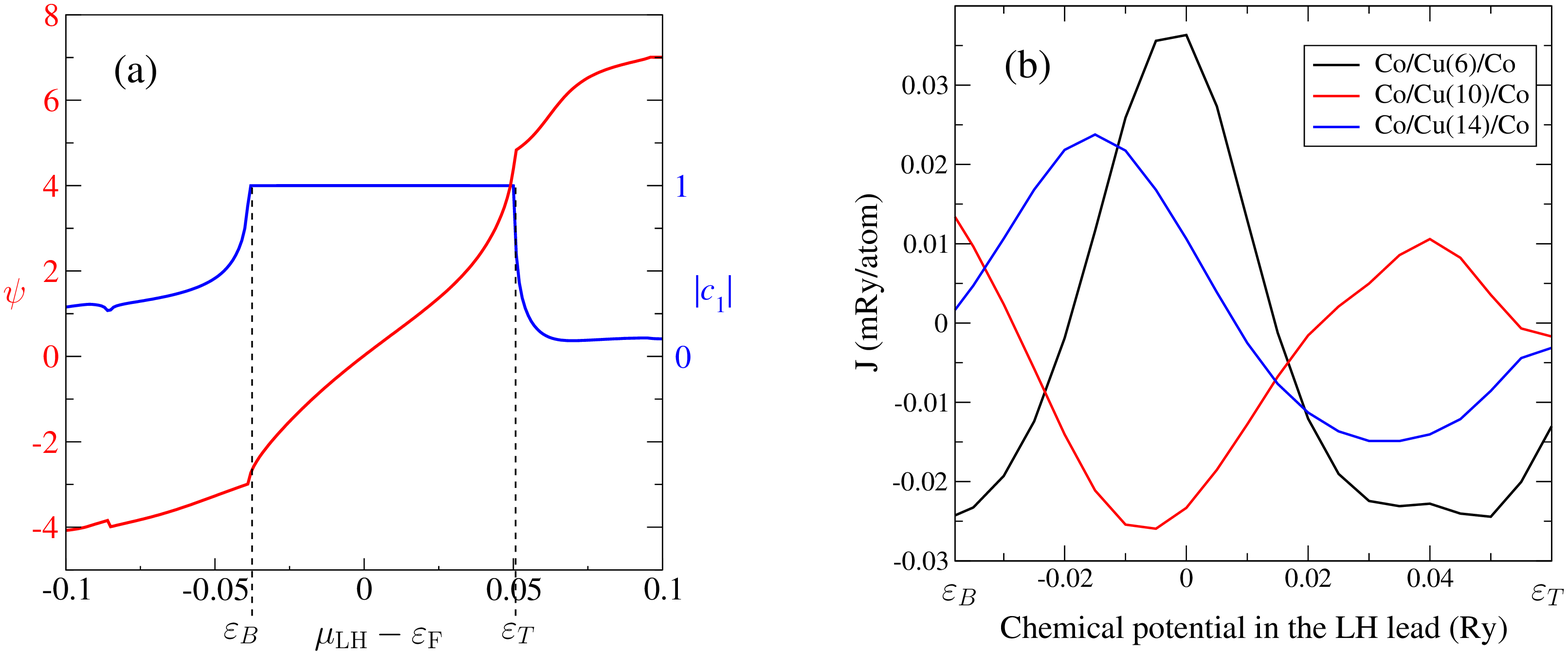}
\end{center}
\vspace{-0.2in}
\caption{\footnotesize (a) The phase (red, left axis) and modulus (blue, right axis) of the Fourier coefficient $c_1$ for FCC Co/Cu/Co (001) (minority spin P configuration) at the neck $\bkp$-point, as a function of the chemical potential ($\mu_{\rm LH}$) in the left-hand lead.\\
(b) The calculated IEC of FCC Co/Cu/Co (100), for 6 (black), 10 (red) and 14 (blue) Cu spacer layers, as a function of the chemical potential in the left-hand lead.  The Co minority HG bottom is at $\veps_B \approx-0.038$ and the top at $\veps_T \approx 0.05$ Ry.}
\label{fig:fig45}
\end{figure}

Luckily there are a number realistic systems where the IEC is dominated by a quantum well state localised in a HG.
Here we focus on one, FCC Co/Cu/Co grown in the $(001)$ direction and in our calculations we use realistic s, p, and d tight-binding  parameters determined from the {\em ab-initio} band structures as reported previously in Ref.\cite{Mathon97}.
There it was shown that Cu has a single Fermi-surface sheet which has two extrema, one at $\bkp=\mathbf{0}$ (the belly) and another at $\bkp\approx(0.8\pi,0.8\pi)$ (the neck).
The contribution to the IEC from the belly extrema is about $400$ times smaller than that of the neck (because there is good matching between Cu and both majority and minority Co bands at the Fermi-level) so is irrelevant to this calculation.
However at the neck, the sp-like Cu band at the Fermi level has no counterpart for the minority spin in Co since it falls into a HG. 
For the Fermi-level set at $\veps_F=0$, the top and bottom of the HG occur occur at $\veps_T \approx 0.05$ Ry and $\veps_B \approx -0.038$ Ry.
If the Fermi level of the lead lies in this regime, our calculations show that the first Fourier coefficient for $\Omega_{\rm P}^{\downarrow}$ has $|c_1|=1$ and the phase change over the HG is $\Delta \psi \approx 2\pi + 0.8$ radians as predicted (see Fig.~\ref{fig:fig45}(a)). 
The fact that it exceeds $2\pi$ is, as in the 2-band model, most likely due to the presence of other decaying Cu states.
Since there is a good match between the Cu and Co majority
spin bands at the Fermi level, the other three IEC components are small and so the exchange coupling is dominated by $\Omega_{\text P}^{\downarrow}$.  

To check the predicted phase dependence of the IEC, we performed a full numerical computation directly from equation~(\ref{eq:IEC}) (as opposed to the stationary phase approximation).
Figure~\ref{fig:fig45}(b) shows our results for Co/Cu($n$)/Co for $n=6$ (black), $n=10$ (red) and $n=14$ (blue) Cu spacer layers, as a function of the chemical potential in the HG of the left lead. We observe that in each case as the chemical potential moves across the gap there is a phase change of just over $2\pi$ in agreement with the theory developed here.
So a change of approximately 40 mRy (20 mRy) in the potential of one (both) lead(s) is guaranteed to give a sign reversal of the IEC.

Note that a large value for $\psi_s^{\prime}=d\psi_s/d\veps$ has been reported previously \cite{Castro96,Mathon97}.
However these authors attributed this to a band-edge effect observed in the parabolic band model \cite{Mathon92}, missing the multi-orbital effect described here.  
Their primary interest was in the impact of $\psi_s^{\prime}$ on the temperature dependence of the IEC which, as can be seen from equation~(\ref{eq:SPA}), is strong if $\psi_s^{\prime}$ is large.
Such a strong temperature dependence has indeed been measured in several IEC systems where the quantum well spacer state sits in a HG of the lead \cite{Celinski95,Persat97}.
Indeed Maat {\it et. al.} \cite{Maat04} measured a value for $\psi_1^{\prime}$ in almost exact agreement with that predicted by the calculation of  Ref.\cite{Castro96}, providing strong evidence for the existence of the effect described here.  

Even more direct evidence was obtained by Ebels {\it et. al.} \cite{Ebels98} who observed phase changes of up to $2\pi$ in the IEC of Co/Ru/Co and Co/Cu/Co when one of the FM leads is doped with small amounts (up to 8\%) of Ag, Au, Cu and Ru.
However once again these authors incorrectly interpreted their results in terms of parabolic-band type models, so did not appreciate the role played by the HG.

In conclusion.  We have shown that realistic systems can undergo a $\approx 2\pi$ ($4\pi$) variation of the IEC as the chemical potential in one (both) lead moves across a HG.
This variation is largely topologically protected and independent of the detailed parameters of the system: dependent only on the behaviour of the GF at the HG edge.
In particular for a lead with a narrow HG a very small variation of its potential can lead to a sign reversal of the IEC. This has obvious capacity for application to magnetic switching.

\begin{acknowledgments}
The author wishes to thank George Mathon for useful discussions.
\end{acknowledgments}
%

\bibliography{phase}

\end{document}